\documentclass[letterpaper]{article} 
\usepackage{aaai2026}  
\usepackage{times}  
\usepackage{helvet}  
\usepackage{courier}  
\usepackage[hyphens]{url}  
\usepackage{graphicx} 
\usepackage{natbib}  
\usepackage{caption} 
\usepackage{algorithm}
\usepackage{algorithmic}
\usepackage{booktabs} 
\usepackage{amsmath}

\usepackage{newfloat}
\usepackage{listings}
\DeclareCaptionStyle{ruled}{labelfont=normalfont,labelsep=colon,strut=off} 
\floatstyle{ruled}
\newfloat{listing}{tb}{lst}{}
\floatname{listing}{Listing}
\title{ReFeed: Retrieval Feedback–Guided Dataset Construction for Style-Aware Query Rewriting}

\author {
    Jiyoon Myung\textsuperscript{\rm 1,\rm 2}\thanks{
        This work was conducted independently and does not represent the views of the affiliated organizations.
    },
    Jungki Son\textsuperscript{\rm 1},
    Kyungro Lee\textsuperscript{\rm 1, \rm 3},
    Jihyeon Park\textsuperscript{\rm 1, \rm 4},
    Joohyung Han\textsuperscript{\rm 1, \rm 5}
}
\affiliations {
    \textsuperscript{\rm 1}MODULABS, \textsuperscript{\rm 2}Samsung SDS, \textsuperscript{\rm 3}Dable,  \textsuperscript{\rm 4}MIRI D.I.H Co., Ltd., \textsuperscript{\rm 5}Lomin\\
    \{jiyoon0424, aeolian83, lkr981147, milhaud1201, ddang8jh\}@gmail.com
}

\usepackage{bibentry}

\begin{document}

\maketitle

\begin{abstract}
Retrieval systems often fail when user queries differ stylistically or semantically from the language used in domain documents. Query rewriting has been proposed to bridge this gap, improving retrieval by reformulating user queries into semantically equivalent forms. However, most existing approaches overlook the stylistic characteristics of target documents—their domain-specific phrasing, tone, and structure—which are crucial for matching real-world data distributions. We introduce a retrieval feedback–driven dataset generation framework that automatically identifies failed retrieval cases, leverages large language models to rewrite queries in the style of relevant documents, and verifies improvement through re-retrieval. The resulting corpus of (original, rewritten) query pairs enables the training of rewriter models that are explicitly aware of document style and retrieval feedback. This work highlights a new direction in data-centric information retrieval, emphasizing how feedback loops and document-style alignment can enhance the reasoning and adaptability of RAG systems in real-world, domain-specific contexts.
\end{abstract}

\section{Introduction}
Information retrieval (IR) systems often fail when user queries differ stylistically or semantically from the language used in domain documents.
In many real-world applications—such as technical manuals, customer-service chat logs, and enterprise knowledge bases—users express problems or requests in informal, context-rich ways, whereas documents are written in formal and domain-specific styles.
This stylistic mismatch leads to retrieval failures even when the underlying intent of the query is correct, limiting the reliability of retrieval-augmented generation (RAG) systems and conversational assistants.

Query rewriting has emerged as a promising approach to bridge this gap.
By reformulating user queries into semantically equivalent forms, rewriting helps retrievers locate relevant documents more effectively.
However, most existing rewriting approaches remain semantic rather than stylistic: they focus on preserving meaning while ignoring how information is actually phrased in the target corpus.
As a result, rewritten queries may still deviate from the linguistic and structural patterns found in domain-specific documents, leaving retrieval performance suboptimal.

Recent advances in feedback-driven query rewriting have shown that retrieval signals can guide better reformulation and ranking.  
However, most approaches integrate such feedback only \textit{during} model training, using it to fine-tune rewriters via reinforcement objectives.  
In contrast, we investigate how retrieval outcomes can be leveraged \textit{before} training—to automatically curate high-quality supervision data for style-aware query rewriting.

To address these gaps, we propose \textbf{ReFeed}, a retrieval feedback–guided dataset generation framework for style-aware query rewriting.
Our method automatically identifies failed retrieval cases from existing QA datasets, leverages large language models (LLMs) to rewrite queries in the style of relevant documents, and validates improvement through re-retrieval.
Only rewritten queries that successfully retrieve the correct document are retained, resulting in a high-precision corpus of (original, rewritten) query pairs that naturally encode both retrieval feedback and document-style adaptation.
This corpus can be used for few-shot prompting or supervised fine-tuning (SFT) to train rewriter models that are explicitly aware of document style and retrieval outcomes.

Preliminary downstream experiments suggest that incorporating these examples into few-shot prompts can improve retrieval recall and RAG response accuracy, indicating the potential utility of style-aware rewriting for robust information retrieval.
By emphasizing feedback loops and stylistic alignment, our work lays the groundwork for training future \textbf{style-aware rewriter models}, providing a data-centric foundation for intelligent and adaptive information retrieval systems.

\begin{figure*}[t]
    \centering
    \includegraphics[width=\linewidth]{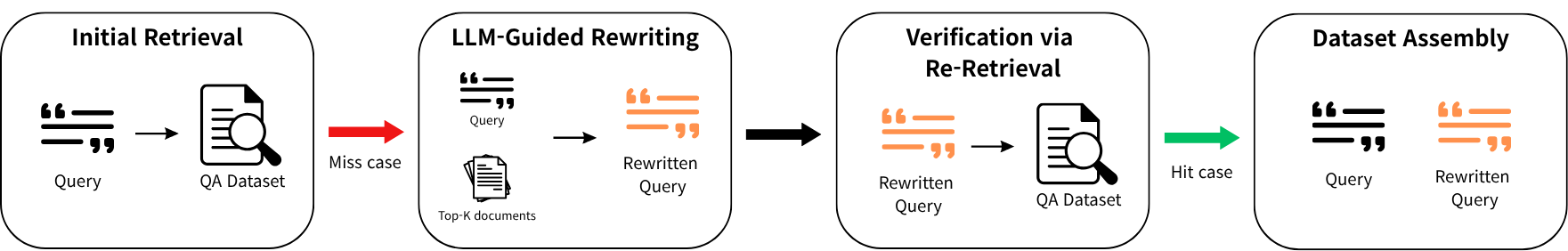}
    \caption{Overview of the retrieval feedback–driven dataset generation framework. 
    Missed queries are rewritten via LLMs to match the style of the correct documents and validated through re-retrieval before being assembled into a style-aware corpus.}
    \label{fig:framework}
\end{figure*}

\section{Related Work}

\paragraph{Query Rewriting for Information Retrieval.}
Query rewriting aims to reformulate user queries into more effective forms for retrieval or generation tasks.  
Recent works have explored explicit rewriting mechanisms optimized for retrieval, including 
structured rewriting frameworks \citep{baek2024crafting} and generative reformulation approaches \citep{dhole2024generative}.  
These studies show that leveraging large language models for rewriting can improve recall and ranking quality, 
but they primarily emphasize semantic reformulation or expansion rather than aligning rewritten queries with the stylistic patterns of target domain documents.

\paragraph{Feedback-Driven Query Reformulation.}
Another line of research integrates retrieval or ranking feedback into the rewriting process.  
For example, \citet{mao2024rafe} leverage reranker-derived ranking signals to fine-tune rewriters, 
and \citet{wang2024maferw} incorporate multi-aspect feedback—including retrieval, generation, and document similarity—to optimize rewriting for RAG systems.  
Similarly, \citet{shi2023replug} augment black-box language models with retrieval feedback to enhance evidence grounding.  
While these methods demonstrate that feedback can guide rewriter optimization, they typically employ reinforcement learning or reward modeling to train models directly.  
In contrast, our framework uses retrieval feedback as a \textit{data curation signal}, constructing reusable (original, rewritten) query pairs that encode both retrieval success and document-style alignment.

\paragraph{Existing Query Rewriting Datasets.}
Large-scale rewriting datasets such as \textbf{CANARD} \citep{elgohary2019canard} and \textbf{QReCC} \citep{anantha2021qrecc} 
support supervised learning for semantic reformulation, providing pairs of conversational and standalone queries.  
However, these datasets do not capture retrieval feedback or stylistic variation.  
Our proposed dataset complements these resources by leveraging retrieval outcomes to ensure effectiveness 
and by explicitly conditioning rewrites on the linguistic style of relevant domain documents.

\section{Methodology}
\label{sec:methodology}

Our goal is to construct a \textit{retrieval feedback--driven dataset} for \textbf{style-aware query rewriting}—a corpus of (original, rewritten) query pairs that reflect both retrieval outcomes and the stylistic characteristics of domain documents.  
The overall framework, illustrated in Figure~\ref{fig:framework}, consists of four stages: (1) Initial Retrieval, (2) LLM-Guided Rewriting, (3) Verification via Re-Retrieval, and (4) Dataset Assembly.

\subsection{Initial Retrieval}

We begin with an existing QA dataset \(D = \{(q_i, a_i)\}\), where each query \(q_i\) is paired with its relevant document or passage \(a_i\).  
Each query is embedded and used to retrieve the top-\(k\) documents from the corpus using a dense retriever (e.g., E5-base or BGE-large).  
If the ground-truth document \(a_i\) is not included in the retrieved top-\(k\), the instance is marked as a \textbf{miss case}.  
These failure cases serve as informative starting points for generating retrieval-aware rewrites.

\subsection{LLM-Guided Rewriting}

For each miss case, we prompt a large language model (LLM) to rewrite the query so that it better matches the language and style of the relevant document.  
The input prompt includes (1) the original query \(q_i\), (2) the top-\(k\) retrieved but incorrect documents \(D_{\text{neg}}\), and (3) the ground-truth document or its summary \(D_{\text{pos}}\).  
A simplified prompt template is shown below:

\begin{quote}
\small
\texttt{Your goal is to refine the given query so that it successfully retrieves the correct document among multiple embedded candidates.}\\
\texttt{}\\
\texttt{Original query: [q\_i]}\\
\texttt{Correct document: [D\_pos]}\\
\texttt{Retrieved but irrelevant documents:}\\
\texttt{[D\_neg\_1 ... D\_neg\_k]}\\
\texttt{}\\
\texttt{Rewrite the query to better match the style and phrasing of the correct document, so that it can be retrieved within the top-k results.}\\
\texttt{}\\
\texttt{Do not copy or use any specific information from the correct document.}\\
\texttt{Revise the query only based on the content and intent of the original query itself.}
\end{quote}

The LLM then produces one or more rewritten queries \(q'_i\).  
By conditioning on both negative and positive contexts, the model implicitly learns to align the query with the target document’s linguistic and stylistic properties rather than performing purely semantic paraphrasing.

\subsection{Verification via Re-Retrieval}

Each rewritten query \(q'_i\) is re-submitted to the same retriever.  
If the ground-truth document \(a_i\) now appears within the top-\(k\) retrieved results, the rewrite is considered a \textbf{hit case}.  
Otherwise, the sample is discarded or optionally re-attempted through iterative rewriting.  
This \textit{re-retrieval} stage ensures that the resulting pairs are empirically validated: every retained rewrite demonstrably improves retrieval performance relative to the original query.

\begin{table*}[h]
\small
\centering
\begin{tabular}{p{0.47\textwidth}p{0.47\textwidth}}
\toprule
\textbf{Original Query} & \textbf{Rewritten Query} \\
\midrule
Certain dogs are bred to help fishermen with what? &  
What specific task are certain dog breeds traditionally used for by fishermen? \\[2pt]
\midrule
What was Gaddafi's nationality? & What was Muammar Gaddafi’s nationality (the country he was a citizen of)? \\[2pt]
\midrule
When copper and tin is mixed, what is made? & What alloy is formed by combining copper and tin? \\[2pt]
\midrule
What was the new name given to Plymouth Dock in 1824? & What was Plymouth Dock renamed to in 1824? \\[2pt]
\midrule
How many did Crazy in Love sell to become one of the greatest selling singles in history? & How many copies did the single ``Crazy in Love'' sell worldwide? \\[2pt]
\bottomrule
\end{tabular}
\caption{Examples of (original, rewritten) query pairs from the constructed dataset.  
The rewriting model dynamically adjusts query specificity and phrasing to better align with the style and terminology of the target corpus—sometimes simplifying overly detailed questions, and other times expanding vague ones for clarity.}
\label{tab:rewrite_examples}
\end{table*}

\subsection{Dataset Assembly}

After verification, we collect all successful pairs \((q_i, q'_i)\) into a new corpus \(D_{\text{rewrite}}\).  
Each entry is stored with metadata such as the retrieval model used, the original ranking position, and the relative improvement in similarity score.  
This corpus forms a \textbf{retrieval-grounded and style-aligned} dataset that can be directly used for training or prompting rewriter models.

\section{Experiments}

\subsection{Setup}

We conduct experiments on the \textbf{SQuAD v1.1} training split to automatically construct a retrieval feedback--driven rewriting dataset.  
Each question is treated as a query \(q_i\), and its corresponding passage as the ground-truth document \(a_i\).  
All passages are embedded using the \textbf{e5-base-v2} model and indexed with \textbf{FAISS} for dense retrieval.  
For each query, we retrieve the top-3 documents based on cosine similarity.

A query is labeled as a \textbf{miss case} if its ground-truth passage does not appear within the top-3 retrieved results.  
Only these miss cases are used in the rewriting stage.

\subsection{Dataset Construction}

For each miss case, we prompt a large language model (\textbf{GPT-5}, temperature = 1.0) following the template described in Section~\ref{sec:methodology}.   
Each rewritten query \(q'_i\) is re-embedded and re-retrieved using the same retriever and index.  
If the ground-truth passage appears within the top-3 results, the pair \((q_i, q'_i)\) is accepted as a \textbf{valid rewrite pair}.

Out of the SQuAD training split (approximately 87k samples), about 16k (18.7\%) of the queries were initially classified as miss cases.  
Among these, the LLM successfully produced valid rewrites for \textbf{67.5\%} of the cases, resulting in a final dataset of roughly \textbf{11,044 verified (original $\leftrightarrow$ rewritten)} pairs.

\paragraph{Qualitative Examples.}
Interestingly, the LLM does not always make queries longer or simpler; instead, it performs \textbf{context-sensitive refinement} based on the retrieval feedback.  
Simple or ambiguous questions tend to be expanded for clarity, while overly complex or redundant ones are simplified to better align with the document style.  
Representative examples from the constructed dataset are shown in Table~\ref{tab:rewrite_examples}.

These examples illustrate that the \textbf{constructed dataset itself} captures a balance between \textit{semantic fidelity} and \textit{retrieval effectiveness}, producing rewritten queries that are stylistically and contextually aligned with the target corpus.   
Consequently, when used to train or prompt rewriter models, this dataset can help them \textbf{internalize retrieval feedback signals} beyond mere paraphrasing, enabling more effective corpus-style reformulation.

\subsection{Few-Shot Validation with Constructed Dataset}

To assess whether the constructed dataset can improve retrieval performance in real-world rewriting scenarios, we conducted a lightweight \textbf{offline few-shot validation} experiment.  
Rather than fine-tuning a rewriter model, we treat the dataset as a retrieval-augmented source of in-context examples.  
Each validation query is embedded, and the five most similar (original, rewritten) pairs from our dataset are retrieved as few-shot demonstrations.  
The following prompt template was used:

\begin{table*}[h]
\small
\centering
\begin{tabular}{p{0.3\textwidth}p{0.3\textwidth}p{0.35\textwidth}}
\toprule
\textbf{Original Query} & \textbf{Rewritten Query} & \textbf{Ground-Truth Rank (Top-\(k\))} \\
\midrule
What is the name of a Time Lord that Doctor Who has fought? & Which Time Lord adversaries has the Doctor fought in the \textit{Doctor Who} series? & From not in top-10 $\rightarrow$ \textbf{Top-2} \\[2pt]
\midrule
In lands attributed to what tribe are found remains of large settlements? & In territories associated with which tribe have archaeologists discovered remains of large settlements? & From Top-8 $\rightarrow$ \textbf{Top-1} \\[2pt]
\midrule
Besides the walk to the church, what else was left out of the day's celebration? & What other parts of the day's celebration were omitted besides the walk to the church? & From Top-5 $\rightarrow$ \textbf{Top-2} \\[2pt]
\midrule
Most imperialism was carried out using which method of transport? & Which mode of transportation was primarily used to carry out imperial expansion? & From not in top-10 $\rightarrow$ \textbf{Top-1} \\
\bottomrule
\end{tabular}
\caption{Few-shot rewriting results using five retrieved example pairs from the constructed dataset.  
Each rewritten query improves retrieval ranking without explicit fine-tuning, indicating that the dataset serves as an effective retrieval-grounded reference for few-shot rewriting.}
\label{tab:fewshot_examples}
\end{table*}

\begin{quote}
\small
\texttt{You are a query rewriter for retrieval.}\\
\texttt{Rewrite the user query into a corpus-style, declarative search query}\\
\texttt{that best matches documentation phrasing.}\\
\texttt{Maintain meaning, avoid hallucinations, prefer canonical terminology,}\\
\texttt{and keep it concise.}\\
\texttt{}\\
\texttt{<Few-shot Examples>}\\
\texttt{Example 1}\\
\texttt{Input: [original query]}\\
\texttt{Rewritten: [refined query]}\\
\texttt{...}\\
\texttt{}\\
\texttt{<Task>}\\
\texttt{User Query: [user query]}\\
\texttt{Rewrite only the query. Respond with the rewritten query text only—no explanations.}
\end{quote}

This setup simulates a realistic offline environment where rewrites are generated on demand using retrieved exemplars, without task-specific training.

\paragraph{Discussion.}
This experiment demonstrates that the constructed dataset is \textit{operationally useful} even without supervised training.  
When used as few-shot exemplars, the verified pairs help the LLM produce retrieval-aligned rewrites that better reflect the linguistic style of documentation.  
Because the SQuAD dataset primarily contains short and well-structured factual questions, rewriting is not beneficial for all queries—many original questions already align with the corpus style.  

Nevertheless, for complex or stylistically divergent queries, such as those involving descriptive expressions, implicit entities, or person-specific phrasing, few-shot rewriting notably improved the ranking of relevant documents.  
Representative examples are shown in Table~\ref{tab:fewshot_examples}.  
Overall, while rewriting may not universally enhance retrieval on simple datasets, these results highlight the potential of retrieval feedback–driven data as a foundation for developing future style-aware rewriter models. 

Moreover, since the rewriting process implicitly adapts to the phrasing style, terminology, and formality of the corpus, the generated queries naturally align with how information is expressed within each domain—suggesting strong applicability to real-world RAG systems.

\section{Conclusion and Future Work}

In this work, we presented \textbf{ReFeed}, a framework for constructing a \textit{retrieval feedback--driven, style-aware query rewriting dataset}.  
By leveraging retrieval outcomes as weak supervision, the framework systematically collects (original, rewritten) query pairs that empirically improve document recall.  
Through LLM-guided rewriting, re-retrieval verification, and curated dataset assembly, the resulting corpus achieves a natural balance between \textit{semantic fidelity} and \textit{retrieval effectiveness}.  

The rewriting process implicitly adapts to the phrasing style, terminology, and formality of the corpus, allowing the generated queries to naturally align with how information is expressed within each domain.  
This corpus-adaptive property makes the approach particularly suitable for retrieval-augmented generation systems, where query formulation must reflect the linguistic characteristics of domain-specific documents.  

Preliminary few-shot validation further demonstrated that these verified examples can improve retrieval performance even without model fine-tuning, highlighting the practical utility of the proposed framework for real-world applications.

\paragraph{Future Work.}
Building on this foundation, our next step is to develop an \textbf{offline style-aware rewriting framework} using this dataset to train small language models (SLMs).  
Unlike the current few-shot setup, the trained model will directly internalize corpus-specific linguistic styles, enabling efficient and robust rewriting for domain retrieval.  
To ensure scalability, we plan to integrate a \textbf{selective rewriting module} that determines whether a user query requires rewriting before retrieval—avoiding unnecessary computation while preserving recall gains.  
We also aim to evaluate this end-to-end system on industrial RAG pipelines, investigating trade-offs between rewrite coverage, retrieval accuracy, and latency.  
Ultimately, we envision this framework as a lightweight yet effective approach for improving information access in retrieval-augmented conversational systems.

\bibliography{main}

\end{document}